\documentclass[twocolumn]{aastex631}

\newcommand{\hb}{\hbox{H$\beta$}}

\newcommand{\gsim}{\lower.5ex\hbox{$\; \buildrel > \over \sim \;$}}
\newcommand{\lsim}{\lower.5ex\hbox{$\; \buildrel < \over \sim \;$}}

\newcommand{\oii}{\hbox{[O\,{\sc ii}]}}
\newcommand{\oiii}{\hbox{[O\,{\sc iii}]}}

\newcommand{\neviii}{\hbox{Ne\,{\sc viii}}}

\newcommand{\feii}{\hbox{Fe\,{\sc ii}}}

\newcommand{\oiv}{\hbox{O\,{\sc iv}}}
\newcommand{\ovi}{\hbox{O\,{\sc vi}}}

\submitjournal{the ApJ Letters}

\shorttitle{HE 0238$-$1904 Superbubble}
\shortauthors{Zhao \& Wang}

\begin{document}


\title{Discovery of a spatially and kinematically resolved 55~kpc-scale superbubble inflated by an intermediate redshift non-BAL quasar}

\author{Qinyuan Zhao}
\affiliation{Department of Astronomy, Xiamen University, Xiamen, Fujian 361005, China}

\author[0000-0003-4874-0369]{Junfeng Wang}
\affiliation{Department of Astronomy, Xiamen University, Xiamen, Fujian 361005, China}



\correspondingauthor{Junfeng Wang}
\email{zqy94070@xmu.edu.cn; jfwang@xmu.edu.cn}

\begin{abstract}

We report on the discovery of a rare case of spatially and kinematically resolved galactic scale outflow at intermediate redshift, based on VLT/MUSE optical integral field spectroscopic observation of the quasar HE 0238$-$1904. This classical non-broad absorption line (non-BAL) quasar at $z=0.631$ remains underexplored in its optical emission lines, though its UV absorption lines are well-studied. We identify a superbubble driven by HE 0238$-$1904 from the emission line morphology, line ratio diagnostics and kinematics, showing a one-sided outflow reaching a projected distance of $R \sim 55$ kpc from the nucleus. The bulk of the ionized gas, with a characteristic mass $M \sim 10^{8}~\rm M_{\odot}$, is blueshifted by $v \approx 700$ km s$^{-1}$ with respect to the quasar systemic velocity.  The outflows detected using absorption line and the emission line are likely stratified components of different spatial scale and velocity in the ionized phase outflow.  Although feedback in HE 0238$-$1904 is taking place on kpc scale, the kinetic power of the outflow at 55 kpc ($\ll 0.1\% L_{bol}$) implies that it is inadequate to regulate effectively the evolution of the host galaxy at this large scale.



\end{abstract}



\section{Introduction} \label{sec:introduction}

Modern galaxy formation theory strongly suggests that there is a fundamental connection between the
supermassive black holes (SMBHs) residing in galaxy centres and the formation and evolution of their host galaxies \citep{Kormendy2013a}. 
Theoretical studies and simulations show that active galactic nuclei (AGN) feedback can provide an explanation for a variety of observations, from the chemical enrichment of the intergalactic medium to the self-regulation of the growth of the SMBHs and of the galactic bulge (e.g. \citealt{Ferrarese2000,Gebhardt2000,Tremaine2002,DiMatteo2005,Hopkins2010}). Powerful outflows driven by AGN have been invoked as one of the main conveyors, so that SMBH activity has a controlling effect on shaping the global properties of the host galaxies \citep{Tabor1993,Silk1998,Springel2005,Croton2006,Hopkins2006,Choi2012,Veilleux2005,Veilleux2020}.

In the past several years, AGN-driven outflows extending to kpc-scales have been resolved in ionized, atomic and molecular gas around both radio-loud (e.g. \citealt{Nesvadba2006,Nesvadba2008,Vayner2021a}) and radio-quiet quasars (e.g. \citealt{Nesvadba2008,Liu2013,Liu2013a}) across low (e.g. \citealt{Feruglio2010,Feruglio2013a,Rauch2013,Cicone2012,Cicone2014}) and high redshift (e.g. \citealt{Alexander2010a,Nesvadba2011,Harrison2012,Harrison2014,Carniani2015,Vayner2021}). 
These energetic outflows and jets emanating from the AGN may inflate galactic scale bubble-like structure along the minor axis  (largely perpendicularly to the main plane of the galaxy) extending beyond tens of kpc \citep[e.g.,][]{Leung2021}, eventually expanding into the intergalactic medium. This is also referred to as a superbubble.  Originally, superbubbles are powered by the combined explosions of supernova in a cluster of massive stars, with cavities of diameter greater than 100 pc and density lower than that of the surrounding interstellar medium (ISM) \citep{TenorioTagle1988,Rupke2013,Zaninetti2021}. The kinetic energy of the optical line emitting gas may reach several times of $10^{55}$ ergs \citep{Cecil2002}. In the recent literature this definition extends to AGN inflated bubbles. A well-known local example is the nuclear superbubble emerging from the edge-on galaxy NGC 3079, although the central starburst appears sufficient to power the outflow, with contribution from the AGN \citep{Irwin1988,Veilleux1994,Cecil2001}. A spectacular 10 kpc bipolar superbubble dominated by AGN radiation is detected in the radio, optical and X-ray band in the ``Teacup AGN" (SDSSJ 1430+1339) at redshift $z=0.085$ \citep{Lansbury2018}.
So far, 10 kpc-scale optical superbubbles driven by AGN have been found around obscured quasars at low redshift (e.g. \citealt{Greene2012}). Such spatially resolved cases through ionized emission line gas are still scarce.

Roughly 20\% of the quasars show blueshifted broad absorption lines (BALs), implying that radiatively driven high velocity outflows are ubiquitous \citep{Proga2000,Hewett2003}. Previous work find that such massive, sub-relativistic outflows can be very efficient feedback agents, based on ultraviolet (UV) BAL analyses \citep{McCarthy2010,Faucher-Giguere2012,Choi2014,Miller2020,Byun2022a}. A recent study using the large sample of SDSS quasars \citep{Rankine2020} concludes that BALs and non-BALs represent different views of the same underlying quasar population, implying that the outflows in BALs and non-BALs quasars are similar. In this work, we present VLT/MUSE discovery of the powerful ionized gas outflow detected in a non-BAL quasar at $z \sim 0.6$.

HE 0238$-$1904 is a typical non-BAL quasar with a central black hole of 2.4$\times$10$^{10}$ M$_\odot$ \citep{Onken2004}. Previous UV spectroscopy reported the detection of highly ionized collimated outflow in this source \citep{Muzahid2012}. Detailed modeling indicates that the outflow has two ionization phases, where the high-ionization phase carries the bulk of the material \citep{Arav2013}. The absorbing gas is blueshifted from the quasar, and the electron density is $n_{e} = 1.2\times 10^{3.79\pm0.17}$ cm$^{-3}$ measured by $N_{ion}(\oiv^{*})/N_{ion}(\oiv)$ ratio in the absorption troughs. The corresponding mass flux and kinetic luminosity is 40 $M_\odot$ yr$^{-1}$ and 10$^{45}$ erg s$^{-1}$, respectively, where the latter is roughly equal to 1 per cent of the bolometric luminosity \citep{Arav2013}. Hence, this outflow is capable of strong interaction with the host galaxy.  Nevertheless, all these UV absorption line analysis results depend on detailed photoionization modeling, and would benefit from spatially resolved studies. 

This paper is structured as follows. We first describe the observations and data reduction in Section 2. In Section 3 and Section 4, we present the analysis of the spectral data and measurements of the gas kinematics. In Section 5, we discuss the gas kinematics measured by integral field spectroscopy (IFS) and BAL analyses.  We summarize our findings in Section 6. Throughout this paper, we adopt a cosmology with $H_{0}$=70 km s$^{-1}$ Mpc$^{-1}$ , $\Omega_{m} = 0.3$, $\Omega_{\Lambda} = 0.7$. The redshift of HE 0238$-$1904 is adopted as $z=0.631\pm 0.001$, measured using {\hbox{Mg\,{\sc ii}}} $\lambda 2798$ \citep{Wisotzki2000} and consistent with values in recent papers (e.g., $z=0.631$, \citealt{Flesch2015,Neeleman2016}; $z=0.629\pm 0.002$, \citealt{Muzahid2012}).

\section{Observations and Data Reduction} \label{sec:style}

HE 0238$-$1904 was observed by VLT/MUSE between November 2016 and February 2017 under European Southern Observatory (ESO) program ID 096.A-0222(A) (PI: Schaye). The spectra were taken in the optical band (wavelength coverage $\lambda$ $\sim$ 4750 $\mathrm{\AA}$ - 9350 $\mathrm{\AA}$ in the observer's frame), covering 2912 $\mathrm{\AA}$ to 5978 $\mathrm{\AA}$ in the rest frame for our target at $z \sim 0.6$. The total on-source integration time is over 8 hours. The field of view (FOV) approximately covers $1' \times 1'$ with a spatial resolution of $0.8'' \times 0.8''$, and a medium spectral resolution of $R = 3500$. The typical seeing is $0.7'' - 1.2''$, and the airmass ranges between $\sim$ 1.0 and $\sim$ 1.4.

After removing  cosmic rays from the raw data using the L.A. Cosmic procedure \citep{Dokkum2001}, we reduce the raw data using the ESO-MUSE pipeline. The final data cubes have a spatial scale of $0.2\arcsec \times 0.2\arcsec$. 
The estimated angular resolution is $\sim$ 0.8$\arcsec$ based on the full width half maximum (FWHM) of point-spread function (PSF) map, which roughly corresponds to a physical scale of 5.5 kpc at the redshift of $z = 0.631$. We estimate the instrumental PSF using the surface brightness profiles of broad emission lines (BELs), i.e., using a 2D Gaussian to fit the BELs map. 

\section{IFS data analysis} \label{sec:dr}

\subsection{Multicomponent Gaussian fitting}  \label{sec:dr}
In order to understand the dynamics and the main properties of the ionized outflows, we perform a kinematical analysis on the forbidden lines. We remove the broad emission line from the quasar nucleus, which is represented by double Gaussians in the spatial regions where the narrow emission lines are negligible. With the quasar contribution removed, we scrutinize the spatially resolved emission lines following the method described in \citet{Zhao2021}.

We perform a two-step spectral fit to delineate the gas kinematics. First, we extract the spectrum in each spaxel, subtract the continuum using interpolation method from wavelength on two sides of the \oiii\ line, where the continuum is free of any line emission and artifacts.  We use the \feii\ template from \citet{Tsuzuki2006} to subtract the \feii\ emission. Secondly, we assume that the \oiii\ doublet is originating from the same upper level, and the intensity ratio is $I(5007)/I(4959)\sim$3. The lines are fitted with the same central velocity and velocity dispersion. The profile of \oiii\ $\lambda$5007 $\mathrm{\AA}$ emission line is generally complex. So \oiii\ $\lambda\lambda$4959,5007 doublet are fitted with a combination of multiple Gaussians by minimizing $\chi^2$ using the Python package {\tt MPFIT}. We fit the \oiii\ profiles to no more than three Gaussians, following \citet{Liu2014}.

The \oiii\ nebulae surrounding HE 0238$-$1904 is spatially resolved by our IFS observation. The \oiii\ map of HE 0238$-$1904 are shown in Figure~\ref{fig:total}a, where the false color is used to represent the intensity on a logarithmic scale. The surface brightness sensitivity (rms noise) of our \oiii\ maps is approximately $\sigma \sim 3 \times 10^{-19}$ erg s$^{-1} $cm$^{-2}$ arcsec$^{-2}$. We use a 1.5$\sigma$ threshold to create this map.

\subsection{Non-Parametric measurements}  \label{sec:dr}

We obtain non-parametric measurements of the emission line profiles following the method described in \citet{Liu2013}. These include:

(i) zeroth-moment map: intensity map of the \oiii\ $\lambda$5007 $\mathrm{\AA}$ line.

(ii) first-moment map: $v_{med}$, median velocity map.

(iii) second-moment map: line width map, $W_{80}$, the velocity width of the line that encloses 80 per cent of the total flux. This is defined as the difference between the velocities at 10 and 90 per cent of cumulative flux: $W_{80} = v_{90} - v_{10}$. 

(iv) asymmetry, defined as $A = \frac{(v_{90}-v_{\rm med})-(v_{\rm med}-v_{10})}{W_{80}}$. With our definition, a profile with a significantly blueshifted wing has a negative $A$ value, while a symmetric profile has $A = 0$.

\begin{figure*}[htb]
\centering
\gridline{\fig{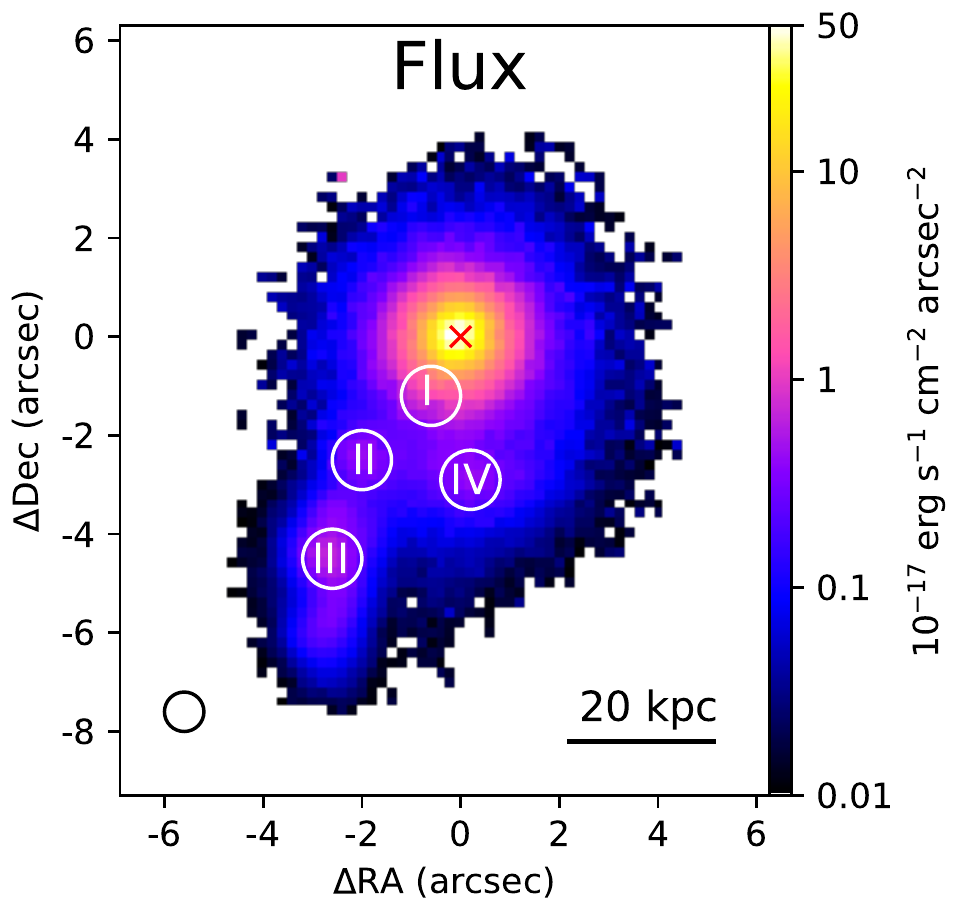}{0.45\textwidth}{a}
\fig{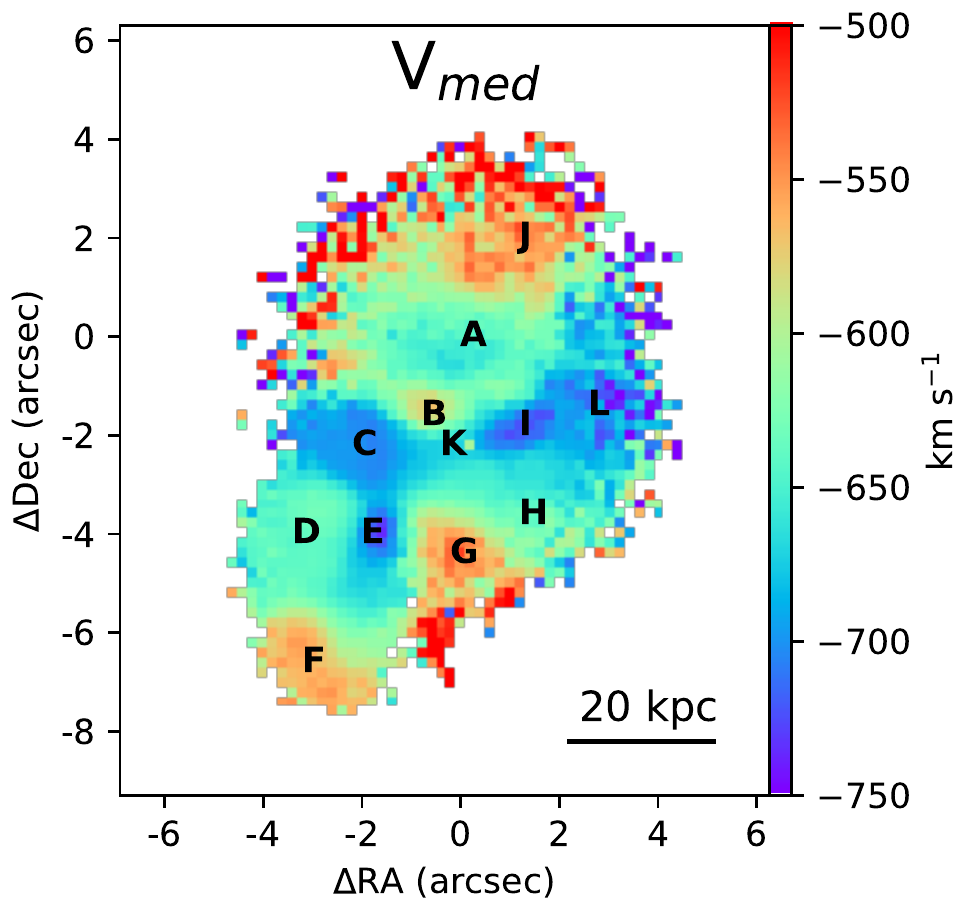}{0.45\textwidth}{b}}
\gridline{\fig{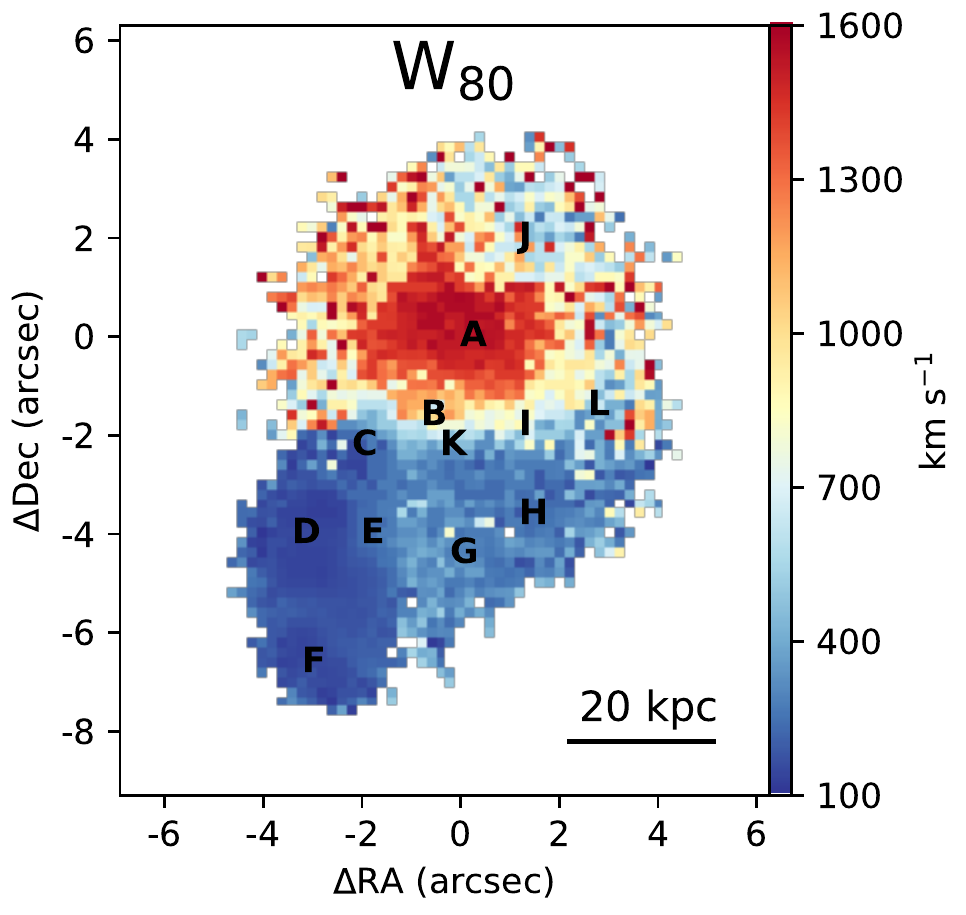}{0.45\textwidth}{c}
\fig{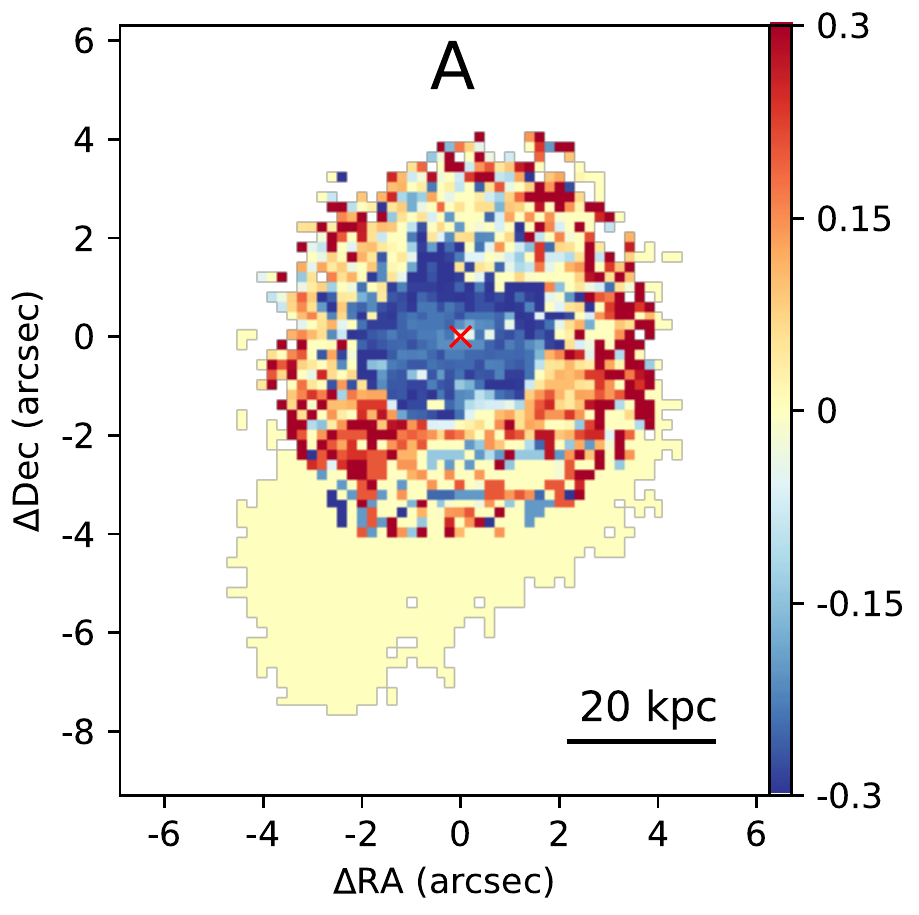}{0.44\textwidth}{d}}

\caption{Non-parametric measurements of HE 0238$-$1904. The maps are (a) flux intensity of \oiii\ (erg s$^{-1}$ cm$^{-2}$ arcsec$^{-2}$); (b) median velocity (km s$^{-1}$); (c) line width ($W_{80}$, km s$^{-1}$); (d) asymmetry ($A$). The maps were obtained by selecting only those spaxels with a S/N of \oiii\ $\lambda$5007 line equal to or higher than 1.5. The red cross marks position of the quasar. PSF (0.8 $\arcsec$) is depicted by the open circle in the lower left of first panel.  
\label{fig:total}}
\end{figure*}

We perform these non-parametric measurements on the best-fitting profiles. Figure~\ref{fig:total} shows these parameters of the ionized gas derived from the fit of the \oiii\ $\lambda$5007 line. The maps are obtained by selecting only pixels with a $S/N \geq 1.5$.

\begin{figure*}[htb]
\centering
\includegraphics[width=0.9\textwidth]{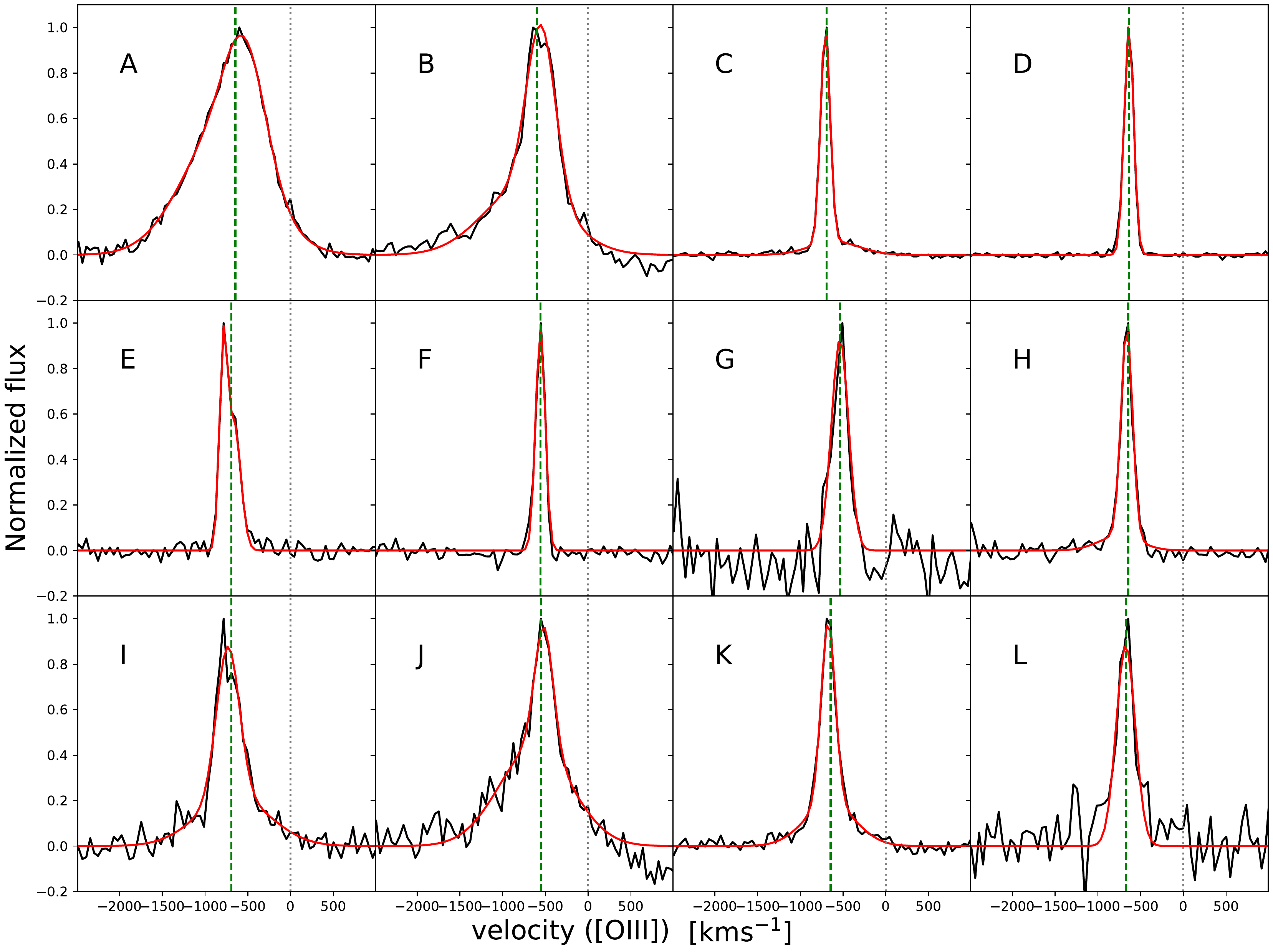}
\caption{We select twelve spatial positions to illustrate the \oiii\ velocity profile therein, including central (``A'', ``B''), and outer regions (``C'' - ``L'') as shown in Figure~\ref{fig:total}. The median and zero velocity is marked by green dotted and grey dashed lines, respectively.
\label{fig:spec}}
\end{figure*}

\section{Properties of the Outflow} 
\label{sec:section_name}

\subsection{Gas Kinematics}  \label{sec:kinematic}

\begin{figure}[h]

\includegraphics[width=0.45\textwidth]{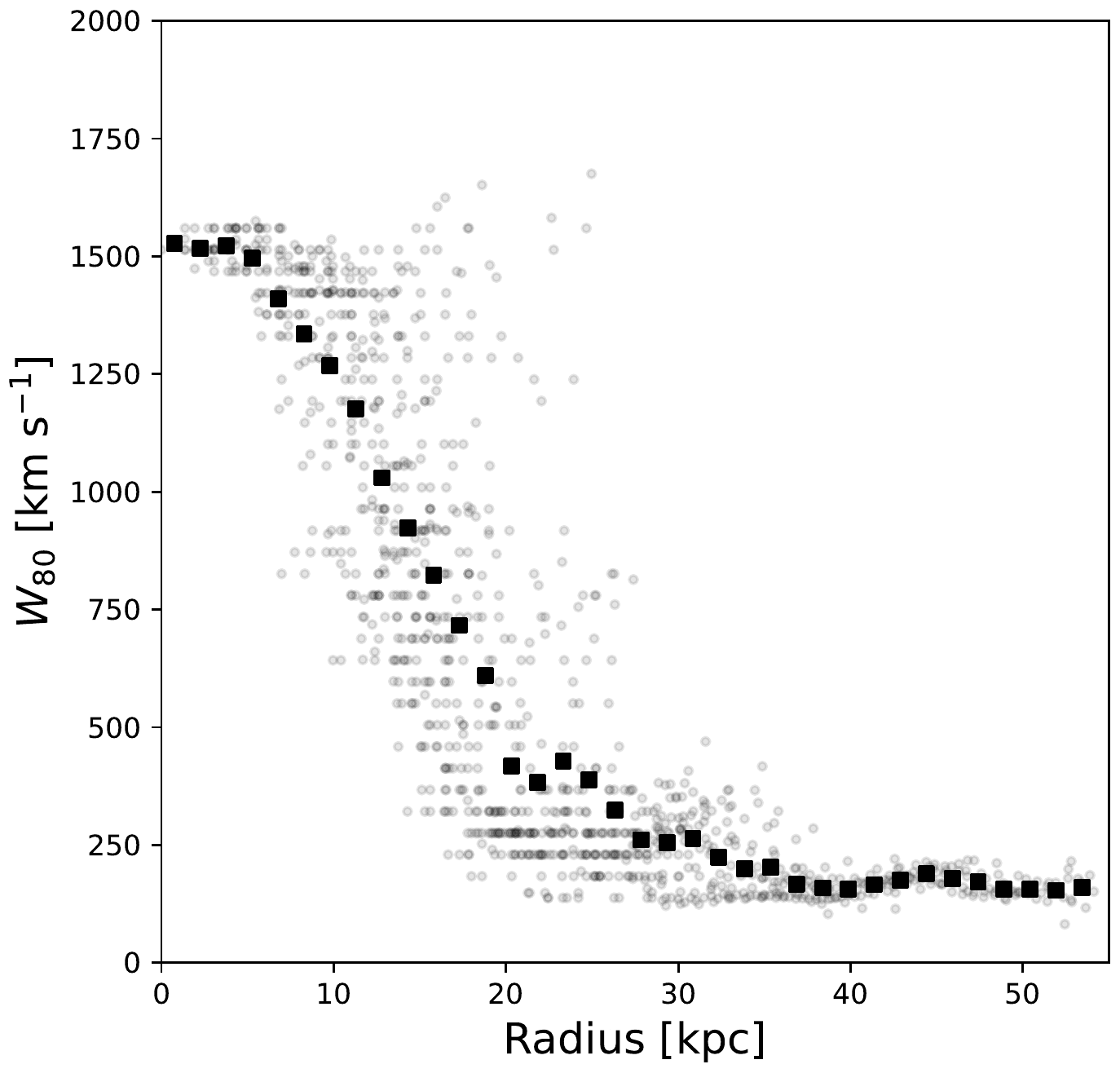}

\caption{The radial dependence of $W_{80}$. Solid squares mark the mean of $W_{80}$ in 1.5 kpc bins using the original data with a S/N of \oiii\ higher than 5. And the grey points shows the original data with a S/N of \oiii\ higher than 5.} 
\label{fig:w80r}
\end{figure}

\begin{figure*}[htb]
\centering
\includegraphics[width=0.9\textwidth]{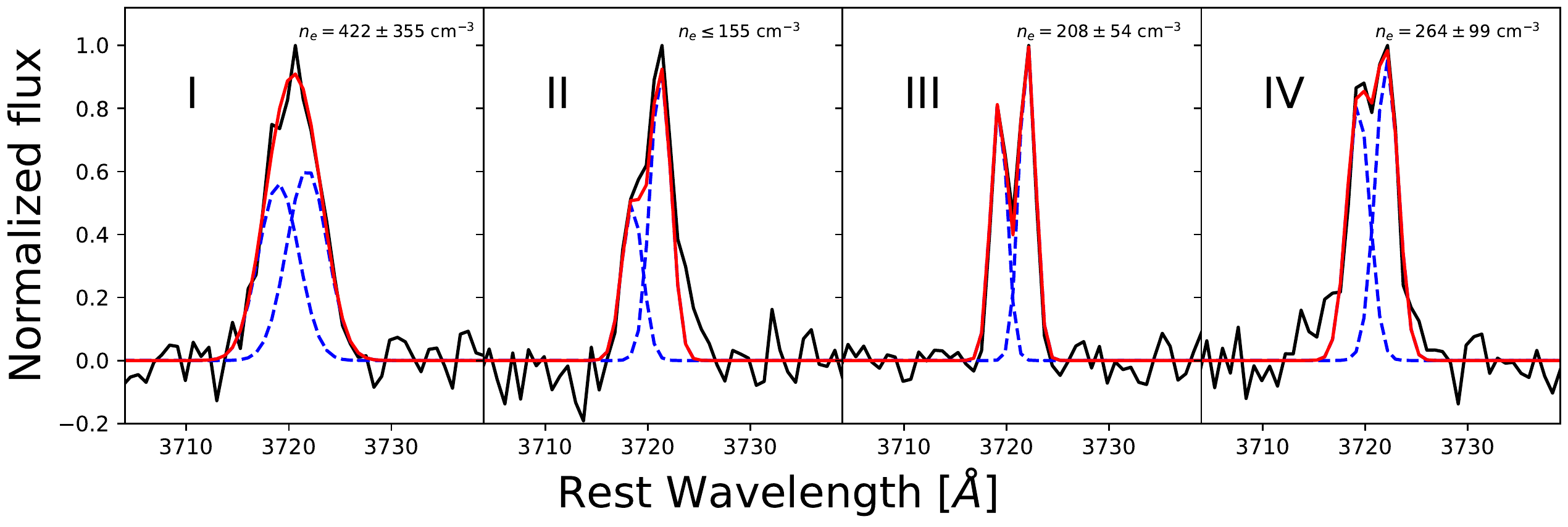}
\caption{We select four spatial positions (Figure~\ref{fig:total}) to present the \oii\ $\lambda\lambda$3737,3729 doublet profiles. The observed spectra and the fitted lines are in black and red, respectively, while the blue dotted lines are the Gaussian components. The electron density in region II is the upper limit in three sigma.
\label{fig:specne}}
\end{figure*}

The forbidden emission line \oiii\ $\lambda\lambda$5007,4969 doublet is adopted as a tracer of ionized outflows on large scale. Figure ~\ref{fig:total} shows the \oiii\ intensity, velocity, velocity dispersion, and asymmetry map. Our IFS data confirmed that \oiii \ $\lambda$5007 lines is spatially and kinematically resolved. In the following, we discuss the velocity, velocity dispersion and asymmetry of the ionized gas and compare them with previous works (e.g. \citealt{Arav2013,Liu2013,Carniani2015}).

The velocity field of the \oiii\ nebulae is remarkably well organized (Figure~\ref{fig:total}b). Only the blue-side outflows is detected, whereas the red side is probably obscured by the host galaxy along the line of sight. For this reason, the \oiii\ line profile is asymmetric, with a prominent blueshifted wing (Figure~\ref{fig:total}d and Figure~\ref{fig:spec}). 

We find the maximum $W_{80}$ value $\sim$ 1600 km s$^{-1}$ in the center, which is comparable to that of known quasar outflows (e.g. \citealt{Liu2013,Carniani2015,Zakamska2016,Kubo2022}), but considerably larger than usual narrow lines in quasar \citep{Lonsdale1993}. The $W_{80}$ value lowers to $\sim$ 200 km s$^{-1}$ in the outer region. The asymmetry map shows regions with heavy blueshifted wings ($<$ 0) that are spatially associated with high-velocity dispersion ($\sim$ 1600 km s$^{-1}$, see Figure~\ref{fig:total}c).

In Figure~\ref{fig:total}(d), we show the map of asymmetry parameter $A$. The asymmetry parameter $A$ is uniformly negative in the bright central part of HE 0238$-$1904, indicating heavily blueshifted wings in the line profiles. This is the tell-tale signature of an outflow which may be proceeding in a symmetric fashion but whose redshifted part is obscured by the material in the host galaxy or near the nucleus \citep{Whittle1985}. In the fainter outer region where the peak S/N of even the brightest emission line \oiii\ is just a few, typically one Gaussian component is sufficient to fit the line profile. Therefore the asymmetry parameter tends to be at zero values ($A = 0$). 

Motivated by the different kinematic components apparent in the velocity map, we extract and fit the \oiii\ lines in 12 regions. Figure~\ref{fig:total}b and Figure~\ref{fig:spec} show the position and spectra of these extraction regions, referred to as A-L. Region A and B correspond to the position of the quasar itself and its immediate vicinity, while C-L correspond to the outer regions. The spectral fits reveal that the width of line profile in center is broader than that of the outer regions. The most significant asymmetry in the line profile is present in the center.
Consequently, the smooth morphology of \oiii\ nebulae, the velocity, the high velocity dispersions of the gas and the blue-shifted asymmetry map all suggest that we have detected ionized outflowing gas in HE 0238$-$1904. 

In Figure~\ref{fig:w80r}, we show values of $W_{80}$ in all spaxels as a function of projected distance from the centre. The radial profiles of $W_{80}$ are almost flat at projected distances $R \lesssim 8$ kpc and appear to decrease at larger radii.  This is different from previous results found in other quasars based on long-slit and IFS observations, which reported flat $W_{80}$ profiles \citep{Greene2011,Liu2013}. One of the concern with the measured decline in $W_{80}$ is that the broad component can no longer be identified in the outer parts, and the $W_{80}$ measurement is due to the narrow component, hence the declining $W_{80}$. In this case, $W_{80}$ parameter would be almost constant across the nebulae in most cases, perhaps declining slightly towards the outer parts. This is not consistent with our $W_{80}$ profile. The possible origin of the rapid decline in $W_{80}$ is discussed in Section ~\ref{sec:origin}.

\subsection{The spatially resolved blobs}  \label{sec:eelr}

The ionized gas extends to the southeast of the nucleus (see Figure~\ref{fig:total}), where three blobs (region II-IV,  Figure~\ref{fig:total}a) and a nuclear region (region I) are present. Both the velocity, velocity dispersion of region II-IV are similar ($v \sim 700$ km s$^{-1}$, $W_{80} \sim 400$ km s$^{-1}$).

\begin{figure}[h]

\includegraphics[width=0.45\textwidth]{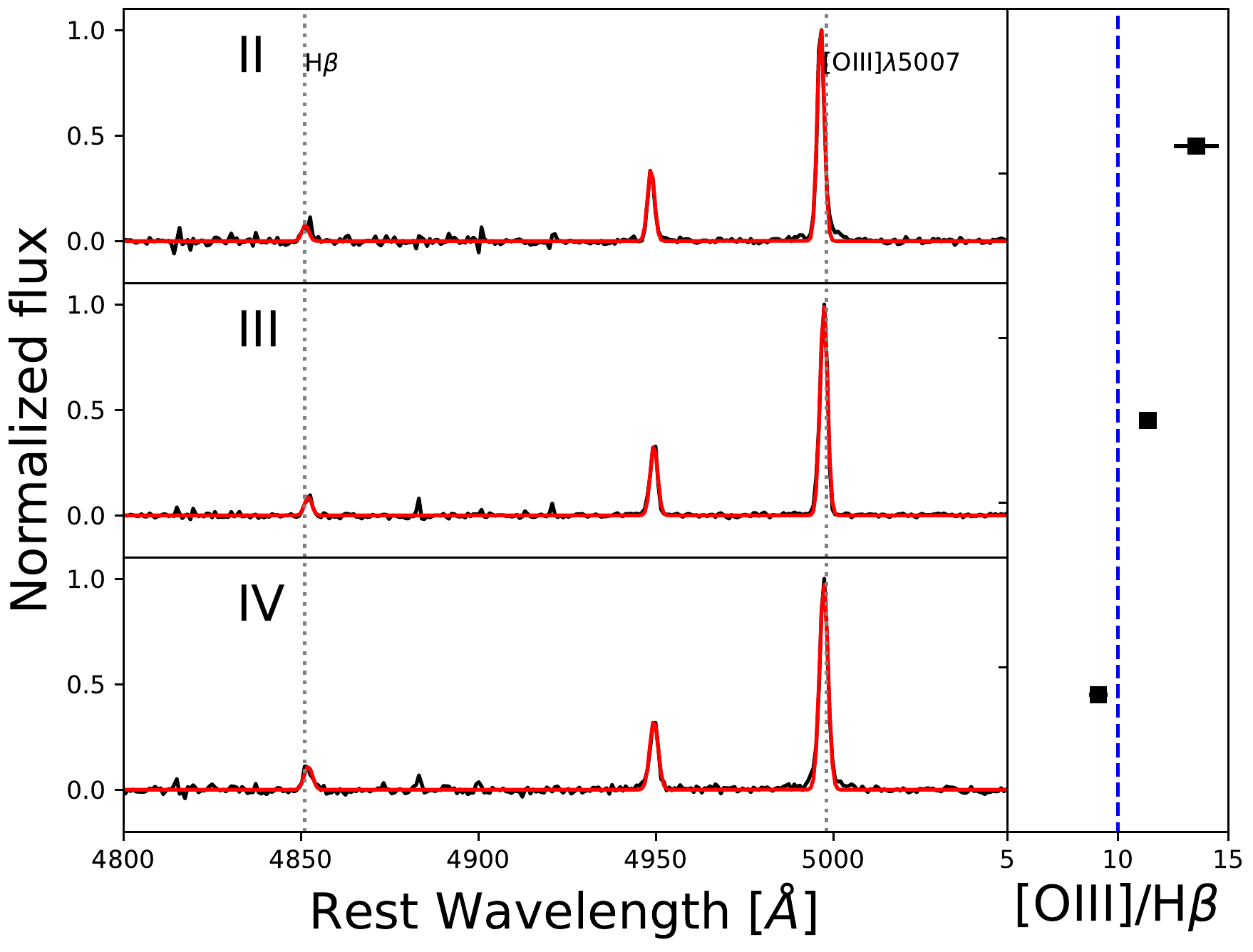}

\caption{Left: Spectra of the \hb\ and \oiii\ doublet at three different regions (II-IV). The red line shows the best fitting, the black line shows the stacked spectra. Right: Line ratio of the \oiii/\hb\ in three regions. Blue dashed lines mark the typical \oiii/\hb\ ratio. The ratio persists at a constant level ($\sim$10) in these three regions. 
} 
\label{fig:ohbr}
\end{figure}

The electron density can be derived from \oii\ $I(3729)/I(3726)$ ratios. For this purpose, we stack the spectra taken from the regions I-IV, and fit the \oii\ doublet emission lines by fixing the kinematics of the two line (Figure~\ref{fig:specne}). Assuming an electron temperature of 10000~K, we estimate the electron density to be $n_{e} = 422\pm355, 208\pm54, 264\pm99$ cm$^{-3}$ in region I, III, and IV, respectively. The electron density in region II is $n_{e} \leq$ 155 cm$^{-3}$, which is the upper limit at three sigma significance. This result is close to that of other quasar outflows (a few 100 cm$^{-3}$; \citealt{Nesvadba2006,Nesvadba2008}). Note that, \oii\ doublet are blended together in the center and difficult to fit, so we stack and fit the spectra from the relatively outer region around the nucleus (region I). Hence, the actual electron density can be higher in the nuclear region.

\section{Discussion} 
\label{sec:discussion}
\subsection{Origin of the ionized gas}  \label{sec:origin}

There are two possibilities for the origin of these ionized gas. The gas can be either a nuclear outflow launched by AGN, or a tidal tail due to galaxy interaction. First, the extended ionized gas is ubiquitous in quasars and generally driven by radiation pressure, so the ionized gas in HE 0238$-$1904 could well be the same. On the other hand, one could suspect that this spatially extended feature with low velocity dispersion (region II-IV) is in fact a tidal tail resulted from interaction with a lower-mass companion. We analyse all the spectra of companion galaxies around HE 0238$-$1904 in the field of view, and the difference between galaxy systemic velocity and these extended emission line gas velocity is found to be at least several hundred km s$^{-1}$. 
This excludes the possibility of ionized gas being tidal tail, because the velocity difference between companion galaxies and tidal tails is generally $\lesssim$ 100 km s$^{-1}$ (e.g. \citealt{Fu2021}).

In addition, the intensity ratios of the emission lines facilitate our analysis on the physical conditions of the ionized gas. Specifically, we use the \oiii/\hb\ ratios to quantify the degree of ionization. To obtain higher S/N ratios, we stack the spectra from the regions II-IV shown in Figure~\ref{fig:total}(a). The \hb\ and \oiii\ spectra and the \oiii/\hb\ ratios in the extended regions (II-IV) are shown in Figure~\ref{fig:ohbr}. 
In this object, \hb\ almost follows the same spatial distribution as that of \oiii\, because \oiii/\hb\ is almost a constant.  The ratio reveals that \oiii/\hb\ is close to 10 in these three regions, implying a high-ionization state in general. Based on the BPT diagram \citep{Baldwin1981}, \oiii/\hb\ $>$ 10 implies AGN dominance. The \oiii/\hb\ profile indicates that the AGN plays a dominant role in the ionization of the large scale gas.

In Figure~\ref{fig:w80r}, we find that the $W_{80}$ parameter is almost flat at projected distances $R \lesssim 8$ kpc  and appear to decrease rapidly at larger radii. Here we discuss the possible origins of the rapid decline of $W_{80}$ with increasing distance from the centre.
This decline of $W_{80}$ implying an apparent narrowing of the line profile is likely due to the fact that the outflow becomes more directional or more collimated within these structures. We thus confirm the previous results based on UV analyses which indicated collimated outflow \citep{Muzahid2012}.

There are various mechanisms capable of establishing an apparently declining $W_{80}$ profile.  First, in the central region the wind expands in all directions from the quasar, whereas at larger distances the opening angle of the outflow is decreased, perhaps because there are low-density regions along which the wind prefers to propagate. Also the outflow in the central region may be experiencing large turbulent motions due to interaction with ISM, but once they escape out of the inner galaxies the flow becomes more organized and mostly radial.  The episodic quasar outbursts may drive a shock wave through the ISM of the galaxy and clear out some of it \citep{Novak2011}.  In any subsequent episodes, the wind suffer less resistance from the ISM and would break out in these directions, producing the large scale bubbles \citep{Faucher-Giguere2012}. Numerical simulations show that this phenomenon is expected for jet-driven winds \citep{Sutherland2007}. We detected three spatially resolved blobs, which are likely part of the rim of the superbubble.  In addition, the outflowing gas is launched and accelerated somewhere close to the quasar, and proceeds ballistically. Thus they eventually slow down as they overcome the potential well of the host galaxy, producing a decline in $W_{80}$.

The ionized outflow traced by \oiii\ emission extends to 50 kpc and beyond, and the $W_{80}$ remains larger than 500 km s$^{-1}$ up to 20 kpc (Figure~\ref{fig:w80r}). Combined with the large-scale morphology and the $W_{80}$ profile, the extended gas we detected closely resembles a superbubble. Future high-quality soft X-ray observations are needed to fully investigate the origin of these ionized gas.

\subsection{Energetics of the Outflow}  \label{sec:dr}

We find that the ionized gas in southwest of the nucleus extends to a projected distance reaching 55 kpc (Figure~\ref{fig:total}). Some characteristic parameters of the ionized gas are further estimated.  Adopting blueshifted velocity of $v_{0} \sim 690$ km s$^{-1}$, and the distance to the galaxy nucleus $R_{0} \sim 55$ kpc.
We estimate a dynamical timescale to be $t \sim 7.8 \times 10^{7}$ yr, i.e. the time required for the gas from the nuclear region to reach such a distance with an average velocity of $v_{0}$.
The mass of gas can typically be estimated using \hb\ luminosity $L_{\hb}$ and electron density $n_{e}$ (e.g \citealt{Liu2013,Harrison2014}).
The total mass of these ionized extended gas can be derived as:
\begin{equation}
 \frac{M_{{\rm gas}}}{2.82\times10^{9}{M_\odot}} = \left(\frac{L_{{\rm H}\beta}}{10^{43}\,{\rm erg\,s}^{-1}}\right) \left(\frac{n_{e}}{100\,{\rm cm}^{-3}}\right)^{-1}
\end{equation}
We take the electron density of $n_{e}$ $\sim$ 400 cm$^{-3}$ in region I, because bulk of the ionized gas is concentrated in central region.
We find $L_{{\rm H}\beta} \sim 3.02 \times 10^{42}$ erg s$^{-1}$, and $M_{\rm gas} \sim 2.1 \times 10^{8}$ $M_\odot$.
Combining with the average velocity of $v \sim$ 690 km s$^{-1}$, the total kinetic energy of the gas can be estimated as:
\begin{equation}
E_{\rm kin} = \frac{1}{2}M_{\rm gas} v_{\rm gas}^{2} = 1.0 \times 10^{57} \rm ~ erg.
\end{equation}
We can also estimate that the mass outflow rate $\dot{M}$ is 2.7 M$_\odot$ yr$^{-1}$, and a kinetic energy rate $\dot{E}_{\rm kin} \sim$ 4.1 $\times 10^{41}$ erg s$^{-1}$. 
The momentum flux of the outflow ($\dot{P}_{outflow}=\dot{M} \times v$) of 1.2 $\times10^{34}$ dynes, or $\log(c \dot{P}_{outflow}/L_\odot)=10.9$. The ratio of the momentum flux of the outflow to the AGN radiation is only 0.002. 

Theoretical modeling predicts that significant feedback requires the kinetic power of the outflow reaches $\sim$ 0.5\%-5\% of the AGN's bolometric luminosity \citep[e.g.,][]{Hopkins2010}. The kinetic luminosity of the outflow detected at 55 kpc is insignificant, only 0.0002\% of its bolometric luminosity  ($L_{bol}=10^{47.2}$ erg s$^{-1}$, \citealt{Arav2013}). This is much lower than the previously estimated $\dot{E}_{\rm kin}/ L_{bol}\sim$1\% from absorption on 3 kpc scale \citep{Arav2013}, or the $\sim 0.05-0.1\%$ estimated for a sample of $z\sim 2.4$ quasars \citep{Carniani2015}. \citet{Carniani2015} fit a a log-linear relation between $\dot{M}$ and $L_{bol}$ for the ionized outflows detected on kpc scale in their quasar sample. Adopting their Equation 10, the expected $\dot{M}\sim 30 M_{\odot}$ yr$^{-1}$ for HE 0238$-$1904 agrees fairly well with the derived $\dot{M}\sim 40 M_{\odot}$ yr$^{-1}$ in \citet{Arav2013}.  In contrast, our estimated mass outflow rate at large scale (55 kpc) in HE 0238$-$1904 is an order of magnitude lower. The kinetic power of the outflow ($\ll 0.1\% L_{bol}$) implies it is no longer an important contributor to feedback at this scale.

\subsection{Comparison with previous UV analysis}  \label{sec:dr}

Using the {\em HST}/COS and {\em FUSE} UV spectra of HE 0238$-$1904, \citet{Muzahid2012} reported detection of outflowing gas in multiple absorption lines \ovi, \neviii, and Mg X. They identified evidence for similar covering factor in several absorption components that kinematically spread over $\sim$1800 km s$^{-1}$, suggesting a collimated outflow. This is consistent with our expectation from radial variation of $W_{80}$. 

Determining the location of the absorption outflow is usually challenging. Nevertheless, assuming a spherical geometry, \citet{Muzahid2012} constrained the radial distance of the absorbing gas to be $R\sim 90$ pc based on the photoionization modeling result.  The detailed UV absorption line analysis by \citet{Arav2013} was able to derive robustly a distance of $R \sim$3 kpc from the outflow to the nucleus using absorption troughs of \oiv, and \oiv*. As revealed by the MUSE data, the projected distance of the outflow from the AGN is $R \sim 55$ kpc for the blueshifted side. This is significantly further from the previous locations of the absorbing gas. In addition, both electron density and the kinematics of these outflows determined from IFS spectroscopy of emission lines are different with those derived from previous UV absorption line analyses \citep[e.g.,][]{Arav2013}. The velocity of the absorption line ($\sim$5000 km s$^{-1}$) is much higher than the velocity of the emission line ($\sim$700 km s$^{-1}$). This strongly indicates that the outflows detected using absorption line and the emission line are clearly not the same component, but likely stratified components of different spatial scale and velocity in the ionized phase outflow.

\section{summary}
In this paper, we present VLT/MUSE IFS observations of a non-BAL quasar HE 0238$-$1904 at redshift $z = 0.631$. \oiii\ emssion lines, are characterised by large line width and prominent blue wings in center, indicative of fast outflows accelerated by the powerful AGN. We summarise our results below:

(1) For the first time, we identify a spatially and kinematically resolved superbubble driven by AGN at intermediate redshift. From the emission line map, ionizing structure and kinematics, we identify a spatially resolved superbubble surrounding HE 0238$-$1904, showing a one-sided structure reaching a projected distance of $R \sim 55$ kpc from the nucleus.  We calculate an electron density $n_{e}$ of a few 100 cm$^{-3}$ for three blobs and the central region. The electron density in center is higher than that in outer regions. The velocity of ionized gas is blueshifted $\sim$ 700 km s$^{-1}$, with a rapidly decline in the radial profile of $W_{80}$. 


(2) We estimate $M_{\rm gas} \sim 2.1 \times 10^{8}~ M_{\odot}$ for the ionized gas mass. The dynamical timescale of the blobs is estimated to be $\sim 78$ Myr, the travel time of the clouds from the centre to reach the observed distance. The resulting mass outflow rate is 2.7 M$_\odot$ yr$^{-1}$.  The kinetic energy carried by the ionized gas is $1.0\times 10^{57}$ erg, with an estimated kinetic energy rate $\dot{E}_{\rm kin} \sim$ 4.1 $\times 10^{41}$ erg s$^{-1}$. Feedback in HE 0238$-$1904 is taking place on kpc scale as previously reported, but the outflow on the 55 kpc scale is inadequate to regulate effectively the evolution of its host galaxy (the kinetic luminosity is only 0.0002\% of the bolometric luminosity).

(3) 
The inferred mass flow rate and kinetic luminosity of the outflow are different from that of previously identified absorption system. The outflows detected in absorption and emission lines are most likely stratified components at different spatial scales and velocities.

\begin{acknowledgments}
We thank the anonymous referee for helpful suggestions that significantly improved our work. We  would like to thank Guilin Liu, Zesen Lin and Luming Sun for the useful discussions. We acknowledge support by the NSFC grants U1831205, 12033004, 12221003, and the science research grants from the China Manned Space Project CMS-CSST-2021-A06 and CMS-CSST-2021-B02. This research has made use of the services of the ESO Science Archive Facility. Based on observations collected at the ESO under program 096.A-0222(A). 
\end{acknowledgments}

%





\bibliography{./ms.bib}{}
\bibliographystyle{aasjournal}



\end{document}